\documentclass{ws-procs975x65}

\begin{document}

\title{Topological and Curvature Effects in a Multi-fermion Interaction Model}

\author{T. INAGAKI and M. HAYASHI}

\address{Department of Physics, Hiroshima University,\\
Higashi-Hiroshima, Hiroshima 739-8526, Japan}

\begin{abstract}
A multi-fermion interaction model is investigated in a compact spacetime with non-trivial topology and in a weakly curved spacetime. Evaluating the effective potential in the leading order of the $1/N$ expansion, we show the phase boundary for a discrete chiral symmetry in an arbitrary dimensions, $2\leq D<4$.
\end{abstract}


\bodymatter

\section{Introduction}\label{inagaki:sec1}
It is believed that a fundamental theory with higher symmetry was realized at the early stage of our universe and the symmetry was broken down to the observed symmetry in the present universe. The mechanism of the symmetry breaking may be found in a dynamics of the fundamental theory. Thus the origin of symmetry breaking is quite important to find a consequence of the fundamental theory in cosmological and astrophysical phenomena.

In QCD the chiral symmetry is dynamically broken according to a non-vanishing expectation value for a composite operator constructed by a quark and an anti-quark field, $\bar{q}q$. The chiral symmetry restoration is theoretically predicted in extreme conditions at the QCD scale. The symmetry restoration at high density may be found in phenomena of dense stars. The heavy ion collision at RIHC and LHC provide experimental data for the symmetry restoration at high temperature.

We can apply the dynamical mechanism to symmetry breaking at GUT era. It is one of candidates to describe symmetry breaking of the fundamental symmetry in GUT. It is natural to expect that the broken symmetry is restored in extreme conditions at GUT scale. Thus we have launched a plan to study the symmetry restoration at high temperature, high density and strong curvature. A topological effect is also interesting before inflationally expansion of our spacetime. In this paper we focus on the topological and the curvature effects. 

A variety of works has been done in a simple four-fermion interaction model. The topological effect has been investigated in the spacetime with one compactified dimension, $S^1$,\cite{Vshivtsev:1995fh} and the torus universe.\cite{Kim:1987db,Song:1990dm,Kim:1994es,IIYF,Abreu:2006pt} It has been found that the finite size effect restore the broken symmetry if we adopt the anti-periodic boundary condition to the fermion fields. On the other hand, the fermion fields which possess the periodic boundary condition contribute to break the symmetry. The curvature effects has been studied in two\cite{Itoyama,BK}, three\cite{EOS}, four\cite{IMO} and arbitrary dimensions.\cite{Inagaki} The broken symmetry is restored if the spacetime curvature is positive and strong enough. However, the symmetry is always broken in a negative curvature spacetime.\cite{Gorbar:1999wa} A combined effect has been also discussed in a weakly curved spacetime,\cite{ELO2} the maximally symmetric spacetime\cite{IMM,ELO} and Einstein space\cite{IIM,Ebert:2008tp}. For a review, see for example Ref.~\refcite{IMO2}.

In these works the four-fermion interaction model is considered to have something essential as a low energy effective model of a fundamental theory. To discuss the model dependence or independence of above results we have to extend the four-fermion interaction model. Here we consider a multi-fermion interaction model\cite{tHooft,Alkofer:1990uh,Osipov,HIT,Ina09} as a simple extension of the four-fermion interaction model and study the contribution from a higher dimensional operator. In Sec. 2 we introduce a multi-fermion interaction model which is considered in this paper. We show an explicit expression of the effective potential in an arbitrary dimensions, $2\leq D<4$. We consider a spacetime $R^{D-1}\otimes S^1$ in Sec. 3. Evaluating the effective potential, we study the topological effect. In Sec. 4 we assume that the spacetime curves slowly and investigate the curvature effect. In Sec. 5 we give some concluding remarks.

\section{Multi-fermion Interaction Model}\label{inagaki:sec2}
As in well-known, the chiral symmetry is dynamically broken
by a simple scalar type four-fermion interaction 
model.\cite{NJL, GN} It is
a useful low energy effective theory of QCD to describe meson
properties. The four-fermion interaction model is also useful 
as a simple toy model in the study of low energy phenomena of 
the strong coupling gauge theory at high energy scale in 
various environments. But there is no reason to neglect higher 
dimensional operators in extreme conditions at the early 
universe. 

In the present paper we extend the model to include scalar type 
multi-fermion interactions,
\begin{equation}
  S=\int d^Dx \sqrt{-g}\left[\sum^N_{l=1}\bar{\psi}_l i\gamma^\mu(x)\nabla_\mu \psi_l 
  + \sum^{n}_{k=1}\frac{G_k}{N^{2k-1}} (\sum^N_{l=1}\bar{\psi}_l\psi_l)^{2k} \right],
\label{Act}
\end{equation}
where index $l$ represents flavors of the fermion field $\psi$,
$N$ is the number of fermion flavors. We neglect the flavor index
below. The multi-fermion interaction is unrenormalizable in 
four spacetime dimensions. The model depends on regularization 
methods. In this paper we adopt the dimensional regularization 
and regards the spacetime dimension, $D$, for the integration of 
internal fermion lines as one of parameters in the effective 
theory.\cite{Naka,He,IKM,JR,IKK}
In QCD it can be fixed to reproduce meson properties.
Here we leave it as an arbitrary parameter to be fixed 
phenomenologically.

The action (\ref{Act}) possesses the discrete chiral symmetry, 
$\bar{\psi}\psi \rightarrow -\bar{\psi}\psi$, and the global 
$SU(N)$ flavor symmetry, 
$\psi \rightarrow e^{i\sum_a \theta_a T_a} \psi$. 
The discrete chiral symmetry prohibits the fermion mass term.
We can adopt the $1/N$ expansion as a non-perturbative approach 
to investigate the dynamical symmetry breaking under the global 
$SU(N)$ symmetry.

For practical calculations it is more convenient to introduce 
the auxiliary fields\cite{Alkofer:1988} and start from the action,
\begin{eqnarray}
S_y=\int d^Dx \sqrt{-g}\left[\bar{\psi}i\gamma^\mu(x)\nabla_\mu \psi 
+ \sum^{n}_{k=1}\frac{NG_k\sigma^{2k} }{(2G_1)^{2k}}
-\frac{N}{2G_1}s\left(\sigma
+\frac{2G_1}{N}\bar{\psi}\psi\right)\right].
\label{Sy1}
\end{eqnarray}
The multi-fermion interactions in the original action are replaced by 
the auxiliary fields $\sigma$ and $s$.
If the auxiliary field $s$ develops a non-vanishing expectation value,
the fermion field acquires a mass term and the chiral symmetry
is eventually broken. 

In this paper we only consider the case $n=2$ for simplicity and 
concentrate on the contribution from the eight-fermion interaction. 
To study the phase structure of the model we calculate the 
expectation value for the auxiliary field, $s$. It is
obtained by observing the minimum of the effective potential. 
It should be noted that the expectation value for the composite 
operator $\bar{\psi}\psi$ is given by the value of the auxiliary field 
$\sigma$ at the minimum of the effective potential.
In the leading order of the $1/N$ expansion we can analytically integrate out the fermion field and get the effective potential,
\begin{equation}
V(s,\sigma)=-\frac{N}{4G_1}(\sigma-s)^2+\frac{1}{4G_1}s^2
-\frac{NG_2}{16G_1^4}\sigma^4
+i \mbox{Tr}\ln\left<x|\left[i\gamma^\mu(x)\nabla_\mu -s
\right]|x\right>.
\label{Eq:pot}
\end{equation}
The trace for the Dirac operator in Eq.(\ref{Eq:pot}) depends on
the spacetime structure. 
The minimum of the effective potential satisfies the gap equation,
\begin{equation}
\left. \frac{\partial V}{\partial \sigma}\right|_s 
 = \left. \frac{\partial V}{\partial s}\right|_{\sigma}
 = 0.
\label{Eq:gap}
\end{equation}
If we consider the four-fermion interaction model, $G_2=0$, 
in the Minkowski spacetime, $R^D$, the gap equation allows a 
nontrivial solution only for a negative $G_1$. The solution 
is give by
\begin{equation}
\sigma=s=m_0\equiv
  \left(\frac{(4\pi)^{D/2}}{\mbox{tr}1\Gamma\left(1-D/2\right)}
  \frac{1}{2G_1}\right)^{1/(D-2)}.
\end{equation}

\begin{figure}[tp]
 \begin{minipage}{0.49\hsize}
  \begin{center}
   \includegraphics[width=60mm]{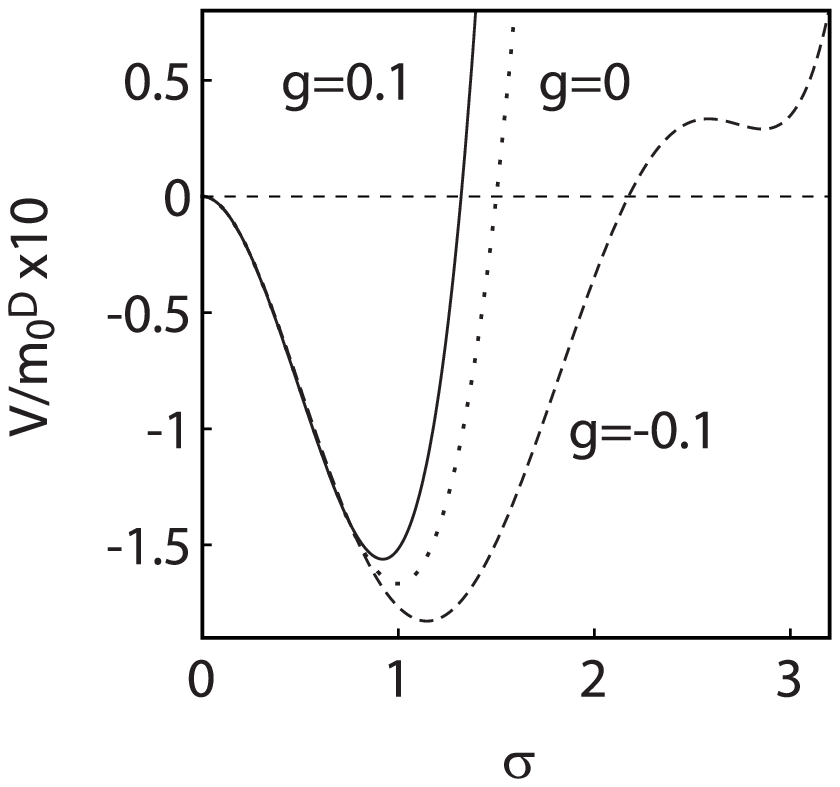}
   \caption{{Behavior of the effective potential in the tree 
dimensional Minkowski spacetime $R^3$. We set $g=0.1, 0$ and $-0.1$ 
and draw the solid, dotted and dashed lines, respectively.}}
   \label{V0}
   \end{center}
  \end{minipage}
  \begin{minipage}{0.02\hsize}
  \end{minipage}
  \begin{minipage}{0.49\hsize}
  \begin{center}
   \includegraphics[width=60mm]{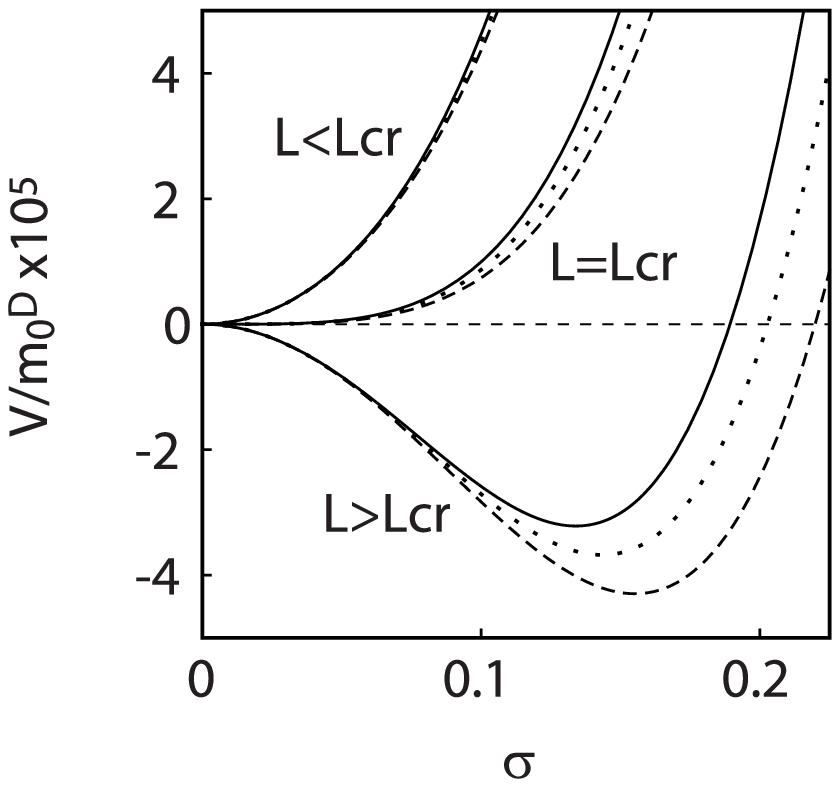}
   \caption{{Behavior of the effective potential on $R^2\otimes S^1$ 
for $\delta_{p,1}=1$,
as $L$ varies. We set $g=0.1, 0$ and $-0.1$ and draw the solid, 
dotted and dashed lines, respectively.}}
   \label{V:An}
   \end{center}
 \end{minipage}
\end{figure}

Before investigating topological and curvature effects, we 
numerically calculate the effective potential in the Minkowski 
spacetime. We are interested in the model where the primordial symmetry is
broken down at low energy scale. Thus we confine ourselves to
a case of a negative $G_1$. In this case the effective potential
 (\ref{Eq:pot}) only depends on $m_0$ and the rate of the coupling 
constants, $G_2 /G_1^3$. We normalize all the mass scales by $m_0$
and set $g=G_2m_0^2/G_1^3$. As is seen in Fig.~\ref{V0}, a positive $g$ 
suppresses the symmetry breaking, while a negative $g$ enhances it. 
A new local minimum appears for a negative $g$.

\section{Phase Structure in a Spacetime with Non-trivial Topology}\label{inagaki:sec3}
One of extreme conditions we have to consider at the early universe is
the topological effect. It is expected that the boundary condition for
matter fields restricts how to compactify the spacetime. In this section
we assume that one of space directions is compactified and investigate
the multi-fermion interaction model on the cylindrical spacetime, 
$R^{D-1}\otimes S^1$. It is a flat spacetime with a non-trivial 
topology.

On $R^{D-1}\otimes S^1$ the effective potential, ($\ref{Eq:pot}$) 
is given by 
\begin{eqnarray}
&&\frac{V(s,\sigma)}{N m_0^D}
=-\frac{\mbox{tr}1}{2(4\pi)^{D/2}}\Gamma\left(1-\frac{D}{2}\right)
  \left(\frac{(\sigma-s)^2}{m_0^2}-\frac{s^2}{m_0^2}
  +\frac{g}{4}\frac{\sigma^4}{m_0^4}\right)
\nonumber \\
&&+\frac{\mbox{tr}1}{2(4\pi)^{(D-1)/2}}\Gamma\left(\frac{1-D}{2}\right)
\frac{1}{Lm_0}\sum_{n=-\infty}^{\infty}\left[
    \left(\frac{(2n+\delta_{p,1})\pi}{Lm_0}\right)^2+\frac{s^2}{m_0^2}\right]^{(D-1)/2},
\label{Eq:potS1}
\end{eqnarray}
where we set $\delta_{p,1}=0$ for the periodic and $\delta_{p,1}=1$ for 
the anti-periodic boundary conditions.
In Fig.~\ref{V:An} the effective potential is shown for 
$\delta_{p,1}=1$, as $L$ varies. It is clearly seen that the broken 
symmetry is restored through the second order phase transition as 
$L$ is decreased. On the other hand only the broken phase is realized 
for the periodic boundary condition.

\begin{figure}[tp]
 \begin{minipage}{0.49\hsize}
  \begin{center}
   \includegraphics[width=60mm]{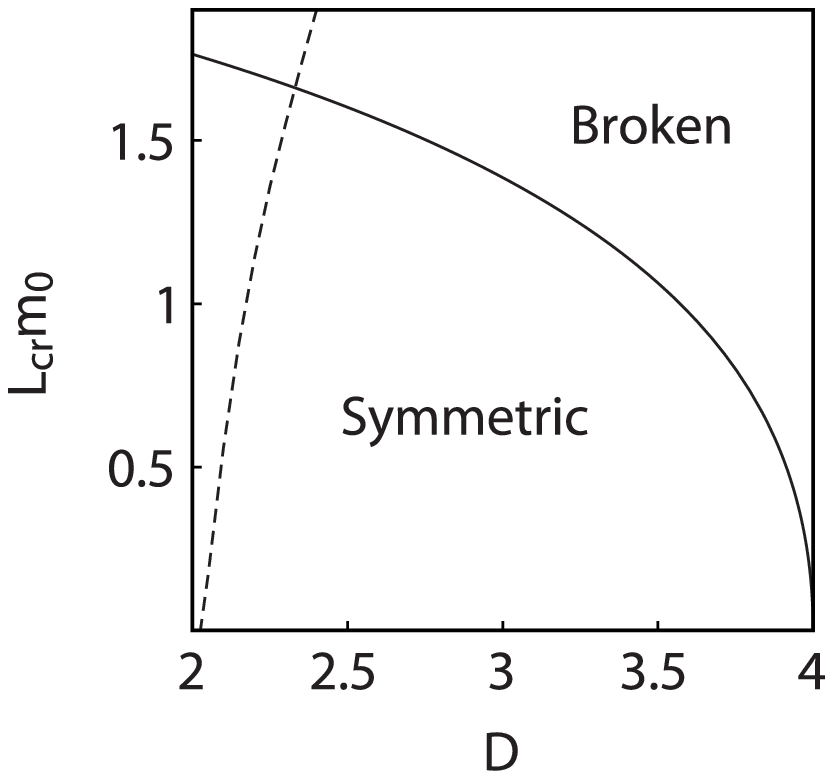}
   \caption{{Critical length. The dashed line shows the length 
to satisfies $V=0$ at the second local minimum for $g=-0.1$.}}
   \label{Cr:An}
   \end{center}
  \end{minipage}
  \begin{minipage}{0.02\hsize}
  \end{minipage}
  \begin{minipage}{0.49\hsize}
  \begin{center}
   \includegraphics[width=60mm]{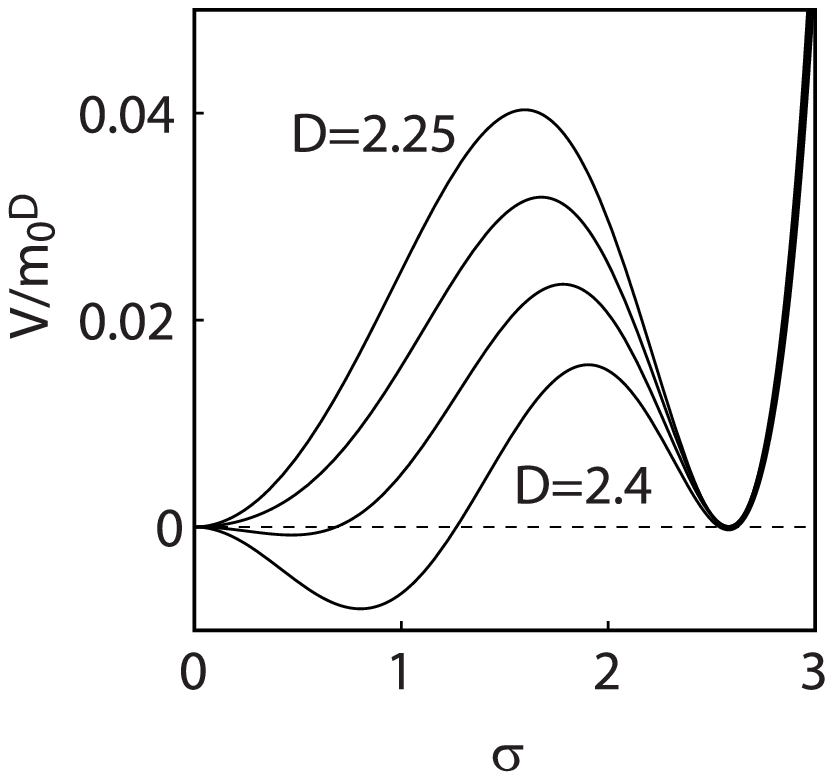}
   \caption{{Behavior of the effective potential on the dashed line in Fig.~\ref{Cr:An} for $D=2.25, 2.3, 2.35$ and $2.4$.}}
   \label{V:An2}
   \end{center}
 \end{minipage}
\end{figure}

To see the situation more precisely we perform a rigorous analysis 
on the critical length $L_{cr}$ for $\delta_{p,1}=1$ and find the 
boundary which divides the symmetric and the broken phases. For a 
non-negative $g$ only the second order phase transition is realized. 
In this case we can find the explicit expression for the critical 
length,\cite{IMO2}
\begin{equation}
L_{cr}m_0
=2\pi\left[\frac{2\Gamma((3-D)/2)}{\sqrt{\pi}\Gamma(1-D/2)}
  (2^{3-D}-1)\zeta(3-D)\right]^{1/(D-2)} .
\end{equation}
It is illustrated by the solid line in Fig.~\ref{Cr:An}. The second
local minimum appears for a negative $g$. The dashed line in 
Fig.~\ref{Cr:An} shows the length $L$ where the effective potential 
satisfies $V=0$ at the second local minimum for $g=-0.1$. 
We plot the effective potential on this dashed line in 
Fig.~\ref{V:An2}. The broken symmetry is restored below both the 
solid and the dashed lines on $D-g$ plane for $g=-0.1$.

\section{Phase Structure in a Weakly Curved Spacetime}\label{inagaki:sec4}
At the early universe the symmetry breaking may be induced under the 
influence of the spacetime curvature. In this section we assume that 
the spacetime curved slowly and keep only terms independent of the
curvature $R$ and terms linear in $R$. We discuss the curvature 
induced phase transition by observing the minimum of the effective 
potential. Following the procedure developed in 
Ref.~\refcite{Inagaki}, we obtain the effective potential up to 
linear in $R$,
\begin{eqnarray}
&&\frac{V(s,\sigma)}{N m_0^D}
=-\frac{\mbox{tr}1}{2(4\pi)^{D/2}}\Gamma\left(1-\frac{D}{2}\right)
  \left(\frac{(\sigma-s)^2}{m_0^2}-\frac{s^2}{m_0^2}
  +\frac{g}{4}\frac{\sigma^4}{m_0^4}\right)
\nonumber \\
&&-\frac{\mbox{tr} 1}{(4\pi)^{D/2}D}
\Gamma\left(1-\frac{D}{2}\right)\frac{|s|^D}{m_0^D}
-\frac{\mbox{tr} 1}{(4\pi)^{D/2}}\frac{R}{24m_0^2}
\Gamma\left(1-\frac{D}{2}\right)\frac{|s|^{D-2}}{m_0^{D-2}}.
\end{eqnarray}

\begin{figure}[tp]
 \begin{minipage}{0.49\hsize}
  \begin{center}
   \includegraphics[width=60mm]{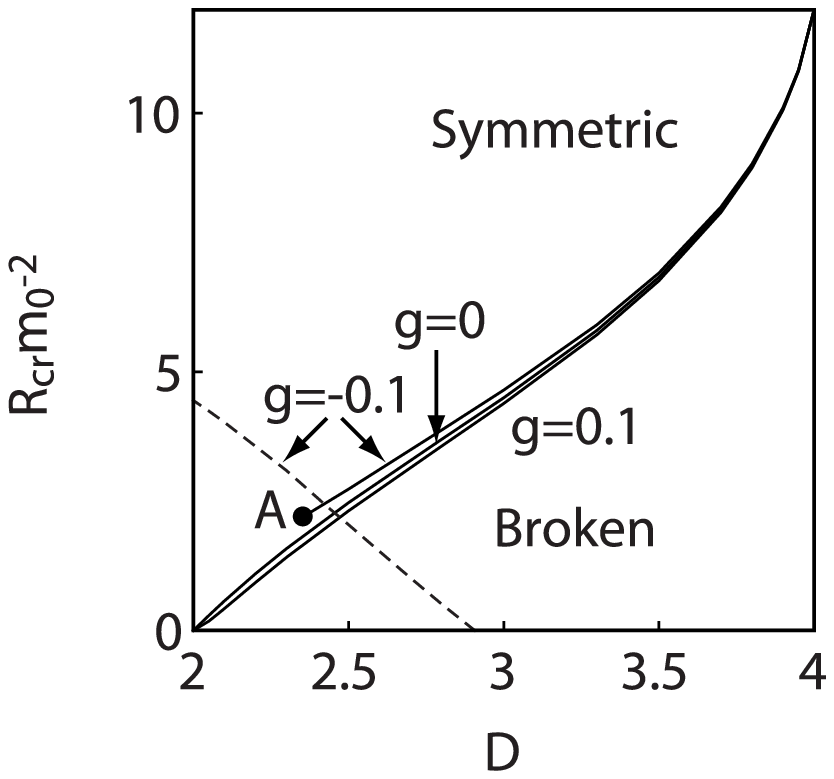}
   \caption{{Critical curvature for $g=-0.1, 0$ and $0.1$. 
The dashed line shows the curvature to 
satisfies $V=0$ at the second local minimum.}}
   \label{cR}
   \end{center}
  \end{minipage}
  \begin{minipage}{0.02\hsize}
  \end{minipage}
  \begin{minipage}{0.49\hsize}
  \begin{center}
   \includegraphics[width=60mm]{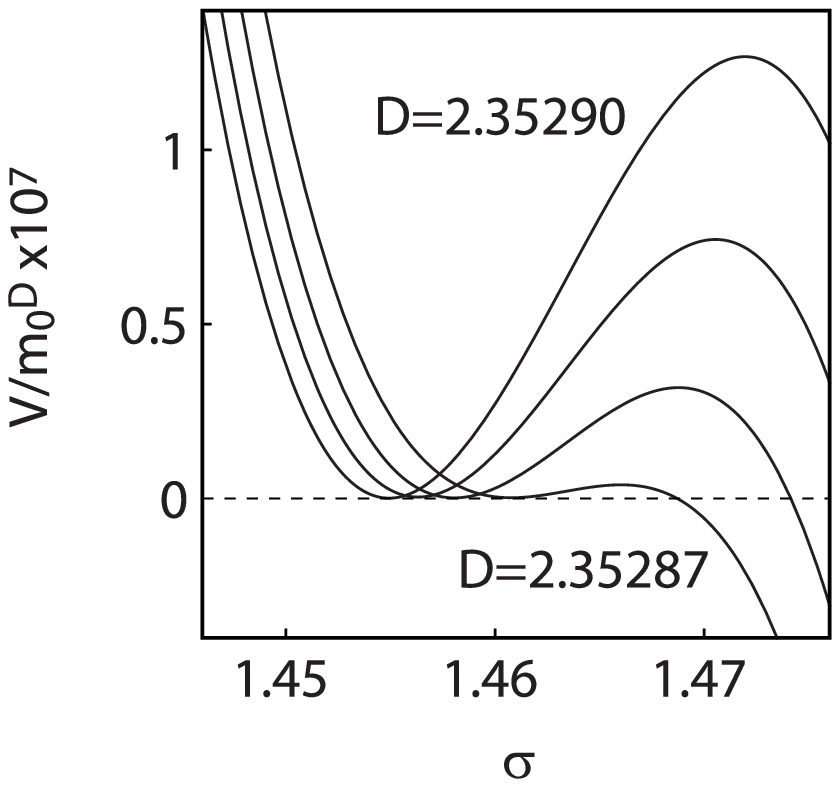}
   \caption{{Behavior of the first local minimum of the effective potential on the solid line for $g=-0.1$ near the point A in Fig.5.}}
   \label{VR}
   \end{center}
 \end{minipage}
\end{figure}

In the four-fermion interaction model, $g=0$, the 
phase transition takes place by varying the curvature $R$. 
The broken symmetry is restored for a large positive curvature 
$R>R_{cr}\geq0$. The phase transition is of the first order for
$2<D<4$. We evaluate the solution of the gap equation 
and find the analytic expression for the critical 
curvature,\cite{Inagaki}
\begin{equation}
\frac{R_{cr}}{m_0^2}
=6(D-2)\left(\frac{(4-D)D}{4}\right)^{(4-D)/(D-2)}.
\end{equation}
It is drawn as a function of the spacetime curvature in Fig.~\ref{cR}.

The critical curvature $R_{cr}$ is modified by the eight-fermion 
interaction. For a positive $g$ we find a smaller $R_{cr}$ except
for two and four dimensions. The dashed line shows the curvature
to satisfy $V=0$ at the second minimum for $g=-0.1$. Thus the
symmetric phase is realized above both the solid and the dashed lines
for $g=-0.1$. In Fig.~\ref{VR} we show the behavior of the effective
potential near the point A on the solid line in Fig.~\ref{cR}. It is
observed that the first local minimum disappears at the point A.
Finally we noted that only the broken phase is realized in a spacetime 
with a negative curvature.\cite{Gorbar:1999wa}

\section{Conclusion}\label{inagaki:sec5}
We have investigated the multi-fermion interaction model under the 
influence of the spacetime topology and curvature. In the practical
calculation the four- and the eight-fermion interactions is studied. 
Evaluating the effective potential in the leading order of the $1/N$ 
expansion, we show the phase structure of the model in an arbitrary 
dimension, $2\leq D <4$.

On $R^{D-1}\otimes S^1$ the finite size 
effect restores the broken chiral symmetry for fermion fields with
the anti-periodic boundary condition. In this case the theory is 
equivalent to the finite temperature field theory.
The phase transition is of the second order for a non-positive 
$g$. We found that the phase boundary is not modified by the 
eight-fermion interaction with a positive $g$. For a 
negative $g$ an additional local minimum contributes to
the phase boundary. A new boundary appears in lower dimensions and
the first order phase transition takes place.

The spacetime curvature also contributes to the phase structure of the
theory. For $2\leq D< 4$ the broken symmetry is restored through the
first order phase transition as increasing the curvature $R$. The 
eight-fermion interaction with a positive $g$ suppresses the chiral 
symmetry breaking and changes the phase boundary. For a negative
$g$ we observe a contribution from an additional local minimum to 
enhance the broken symmetry in lower dimensions.

In the present paper our study is restricted on the analysis of the
phase structure of the theory. We are interested in applying our 
results to critical phenomena at the early universe. 
We will continue our work and hope to report on these problems.

\section*{Acknowledgment}
The authors would like to thank H.~Takata, D.~Kimura, Y.~Kitadono and
 Y.~Mizutani for fruitful discussions.
T.~I. is supported by the Ministry of Education, Science, Sports and
Culture, Grant-in-Aid for Scientific Research (C), No.~18540276, 2009.


\begin{thebibliography}{99}
\bibitem{Vshivtsev:1995fh}
  A.~S.~Vshivtsev, K.~G.~Klimenko and B.~V.~Magnitsky,
  Phys.\ Atom.\ Nucl.\  {\bf 59}, 529 (1996)
  [Yad.\ Fiz.\  {\bf 59}, 557 (1996)].
\bibitem{Kim:1987db}
  S.~K.~Kim, W.~Namgung, K.~S.~Soh and J.~H.~Yee,
  Phys.\ Rev.\  D {\bf 36}, 3172 (1987).
  \bibitem{Song:1990dm}
  D.~Y.~Song and J.~K.~Kim,
  Phys.\ Rev.\  D {\bf 41}, 3165 (1990).
\bibitem{Kim:1994es}
  D.~K.~Kim, Y.~D.~Han and I.~G.~Koh,
  Phys.\ Rev.\  D {\bf 49}, 6943 (1994).
\bibitem{IIYF}
  K.~Ishikawa, T.~Inagaki, K.~Yamamoto and K.~Fukazawa,
  Prog.\ Theor.\ Phys.\  {\bf 99}, 237 (1998).
\bibitem{Abreu:2006pt}
  L.~M.~Abreu, M.~Gomes and A.~J.~da Silva,
  Phys.\ Lett.\  B {\bf 642}, 551 (2006).
\bibitem{Itoyama}
  H.~Itoyama,
  Prog. Theor. Phys. {\bf 64}, 1886 (1980).
\bibitem{BK}
  I.~L.~Buchbinder and E.~N.~Kirillova,
  Int. J. of Mod. Phys. A{\bf 4}, 143 (1989).
\bibitem{EOS}
  E.~Elizalde, S.~D.~Odintsov and Yu.~I.~Shil'nov,
  Mod. Phys. Lett. {\bf A9}, 913 (1994).
\bibitem{IMO}
  T.~Inagaki, T.~Muta and S.~D.~Odintsov,
  Mod. Phys. Lett. {\bf A8}, 2117 (1993).
\bibitem{Inagaki}
  T.~Inagaki,
  Int. J. of Mod. Phys. {\bf A11}, 4561 (1996).
\bibitem{Gorbar:1999wa}
  E.~V.~Gorbar,
  Phys.\ Rev.\  D {\bf 61}, 024013 (2000).
\bibitem{ELO2} E.~Elizalde, S.~Leseduarte and S.~D.~Odintsov,
  Phys. Rev. {\bf D49}, 5551 (1994).
\bibitem{IMM} T.~Inagaki, S.~Mukaigawa,~T. Muta,
  Phys. Rev. {\bf D52}, 4267 (1995).
\bibitem{ELO}
  E.~Elizalde, S.~Leseduarte, S.~D.~Odintsov and Yu.~I.~Shilnov,
  Phys. Rev. {\bf D53}, 1917 (1996).
\bibitem{IIM}
  K.~Ishikawa, T.~Inagaki, T.~Muta,
  Mod. Phys. Lett. {\bf A11}, 939 (1996).
\bibitem{Ebert:2008tp}
  D.~Ebert, K.~G.~Klimenko, A.~V.~Tyukov and V.~C.~Zhukovsky,
  Eur.\ Phys.\ J.\  C {\bf 58}, 57 (2008).
\bibitem{IMO2}
  T.~Inagaki, T.~Muta and S.~D.~Odintsov,
  Prog. Theor. Phys. Suppl. {\bf 127}, 93 (1997).
\bibitem{tHooft}
  G.~'t~Hooft, 
  Phys. Rev. {\bf D14}, 3432 (1976); {\bf D18}, 2199 (E) (1978); 
  Phys. Rep. {\bf 142}, 357 (1986).
\bibitem{Alkofer:1990uh}
  R.~Alkofer and I.~Zahed,
  Phys. Lett.  B {\bf 238}, 149 (1990).
\bibitem{Osipov}
  J.~Moreira, A.~A.~Osipov, B.~Hiller, A.~H.~Blin, J.~Providencia, 
  Annals of Physics {\bf 322}, 2021 (2007);
  A.~A.~Osipov, B.~Hiller, A.~H.~Blin, 
  J. da Providencia, Phys. Lett. {\bf B650}, 262 (2007);
  A.~A.~Osipov, B.~Hiller, J.~Moreira, A.~H.~Blin, 
  J. da Providencia, Phys. Lett. {\bf B646}, 91 (2007).
\bibitem{HIT}
  M.~Hayashi, T.~Inagaki and H.~Takata,
   to be published in Int. J. Mod. Phys. {\bf A},
   arXiv:0812.0900 (hep-ph).
\bibitem{Ina09} T.~Inagaki, {\it The Problems of Modern Cosmology}, 
  ed. P.~M.~Lavrov, (Tomsk Pedagogical University Press, 2009), pp.~214-221.
\bibitem{NJL}
  Y.~Nambu and G.~Jona-Lasinio, 
  Phys. Rev. {\bf 124}, 246 (1961).
\bibitem{GN}
  D.~J.~Gross and A.~Neveu,
  Phys. Rev. {\bf D10}, 3235 (1974).
\bibitem{Naka}
  S.~Krewald, K.~Nakayama, Ann. Phys. {\bf 216}, 201 (1992).
\bibitem{He}
  H.-J.~He, Y.-P.~Kuang, Q.~Wang, Y.-P.~Yi, 
  Phys. Rev. {\bf D45}, 4610 (1992).
\bibitem{IKM}
  T.~Inagaki, T.~Kouno and T.~Muta,
  Int. J. Mod. Phys. {\bf A10}, 2241 (1995).
\bibitem{JR}
  R.~G.~Jafarov, V.~E.~Rochev, Russ. Phys. J. {\bf 49}, 364 (2006).
\bibitem {IKK}
  T.~Inagaki, D.~Kimura and A.~Kvinikhidze, Phys. Rev. {\bf D77}, 
  116004 (2008);
  T.~Fujihara, D.~Kimura, T.~Inagaki and A.~Kvinikhidze, 
  Phys. Rev. {\bf D79}, 096008 (2009).
\bibitem{Alkofer:1988}
  H.~Reinhardt, R.~Alkofer, Phys. Lett. {\bf B207}, 482 (1988).
\end{thebibliography}
\end{document}